\documentclass[twoside,12pt]{article}
\usepackage{graphicx}

\begin{document}

\centerline{\Large\bf Simple Bit-String Model for Lineage Branching}

\bigskip

{\bf P.M.C. de Oliveira$^1$, J.S. S\'a Martins$^1$, D. Stauffer$^2$ and S. 
Moss de Oliveira$^1$}

\bigskip

{\bf Laboratoire de Physique et M\'ecanique des Milieux H\'et\'erog\`enes\par
 \'Ecole Sup\'erieure de Physique et de Chimie Industrielles\par
 10, rue Vauquelin, 75231 Paris Cedex 05, France
}

\bigskip

Permanent addresses:\\

1 Instituto de F\'{\i}sica, Universidade Federal Fluminense, av.  Litor\^anea
s/n, Boa Viagem; Niter\'oi 24210340, RJ, Brazil.\\

2 Institute for Theoretical Physics, Cologne University, D-50923 K\"oln,
Euroland.\\

\centerline{e-mail: pmco@if.uff.br}
\bigskip

\begin{abstract}

        We introduce a population dynamics model, where individual ge-
nomes are represented by bit-strings. Selection is described by death
probabilities which depend on these genomes, and new individuals
continuously replace the ones that die, keeping the population constant. An
offspring has the same genome as its (randomly chosen) parent, except for a
small amount of (also random) mutations. Chance may thus generate a newborn
with a genome that is better than that of its parent, and the newborn will
have a smaller death probability. When this happens, this individual is a
would-be founder of a new lineage. A new lineage is considered created if
its alive descendence grows above a certain previously defined threshold.
The time evolution of populations evolving under these rules is followed by
computer simulations and the probability densities of lineage duration and
size, among others, are computed. These densities show a scale-free
behaviour, in accordance with some conjectures in paleoevolution, and
suggesting a simple mechanism as explanation for the ubiquity of these
power-laws.

\end{abstract}
\medskip

Keywords: Evolution, lineage branching, Monte Carlo simulations

\vfill\newpage

\section{Introduction}

        Biological evolution of species presents some universal behaviour
due to its time-and-size scaleless character (see, for instance,
\cite{Kauffman}). A parallel between this feature and critical phenomena
studied within statistical physics is straightforward, and indeed many
techniques traditionally used by physicists in this field were recently
adopted also to study evolution through simple computer models (see, for
instance, \cite{livro99}). Two of the most important lessons physicists have
learnt from critical phenomena are listed below.

        Lesson 1: One { \bf cannot} take only a small piece (or a small time
interval) of the system under study, including later the rest of the system
as a perturbation. Critical, scale-free systems resist to this approach,
because they are non-linear, the whole is not simply the sum of the parts.
All scales of size (and time) are equally important for the behaviour of the
whole system. A would-be upper bound for size (or lifetime), above which one
can neglect the corresponding effects, does not exist.

        Lesson 2: The specific microscopic (or short term) details of the
system are not definitive to determine the behaviour of the whole system
under a macroscopic (or long term) point of view. In other words, systems
which are completely different in their microscopic constituents (or short
term evolution rules) can present the same critical, macroscopic behaviour.
In particular, some universal critical exponents determine a mathematical
behaviour that is shared by completely distinct systems. Thus, one can
indirectly study some aspects of a complicated real system by observing the
evolution of an artificially invented toy model simulated on the computer.

	There are many evidences for this scale-free behaviour within
biological evolution. Among others, a famous example is the classification
of extinct genera according to their lifetime, a long term study of fossil
data performed by paleontologists John Sepkoski and David Raup
\cite{Sepkoski,Raup1,Raup2}. The frequency distribution they found is
compatible with a power-law decay with exponent 2. The same exponent was
confirmed by at least two distinct theoretical computer models
\cite{Newman,Sole}.

	Branching processes in general also show scale-free behaviour. In
this case, an important class, with exponents multiple of $1/4$, is
ubiquitous. This interesting issue was studied by G.B. West and
collaborators, a recent overview can be found in \cite{West04}. In
particular, by studying blood transport networks, they proposed a model
based on three basic ingredients: a hierarchical branching pattern, where a
vessel bifurcates into smaller vessels and so on; a minimum cut-off size for
the smallest branches, which makes the branching mechanism a finite process;
and a free-energy minimisation constraint. From these three basic
hypotheses, they were able to show the emergence of the exponents $1/4, 1/2,
3/4$, etc \cite{West99,West98,West97}. Of course, not only blood vessel
systems follow this general framework, and the same class of exponents
multiple of $1/4$ were indeed measured within many other contexts.

	A particularly intriguing example is the so-called Kleiber's
empirical law, discovered in 1932. It relates the metabolic energetic power
$P$ of an animal (mammal) with its mass $M$ as $P \sim M^{3/4}$. The
validity of this relation goes down to single isolated mammalian cells and
even its isolated mitochondrian, covering 26 orders of magnitude
\cite{West99}. Also, lifespan increases with $M^{1/4}$ for many organisms,
while heart-rate decreases with $M^{-1/4}$. Thus, the number of heart-beats
during the whole life is invariant for all mammals. Similar scaling
relations and invariant quantities appear at the molecular level as well
\cite{West99}.

	Here, we raise the idea that biological speciation could fit very
well into the general branching process framework described by West. Why
would the idea of universality apply to evolutionary systems is an
interesting and important conceptual question. Some hints towards a possible
answer can be seen in \cite{Doebeli,Geritz,Parisi,PM2001}.

        In the present work, in order to test this possible link between
biological speciation and West's framework, we address such a complicated
problem, namely lineage branching, following the quoted toy model approach.
Our hope is that some of the quantities we can measure could have a parallel
in the real world, in particular the critical exponents. Besides the
computer simulations from which we measure these quantities and their
related critical exponents, we were also able to relate them with each
other. This further analytical treatment yields some scaling relations which
are completely fullfilled by our simulational results. Furthermore, these
relations allow us to predict the unknown values of some exponents from the
knowledge of others, an approach which could be very useful since only one
such exponent was directly measured by fossil data, namely from the Sepkoski
and Raup work. First, we present the model, then the results of our computer
simulations and analytical approaches. Conclusions are at the end.

\vfill\newpage\section{The model}

        Our population is kept constant, with $P$ (typically $10^5$ or $10^6$)
individuals representing a sample of a much larger set. Each individual is
characterised only by its genome, represented here by an array of $g$ bits
(typicaly 32, 64, 128 $\dots$ 2048). Each bit can either be set (1-bit) or not
(0-bit). At the beginning, all bits are zeroed, and all individuals belong to a
single lineage.

        We count the total number $N_i$ of bits set along the genome of
individual $i$: it will {\bf survive} with probability $x^{N_i+1}$, which
decreases exponentially for increasing values of $N_i$, i.e. the larger the
number of 1-bits along the genome, the larger is the death probability of
this particular individual. This is the selection ingredient of our model.
At each time step, a certain fraction $b$ (typically 1\% or 2\%) of
individuals die, each one according to its own death probability, as the
outcome of intralineage competition.

        At each time step, the simulation obtains the value of $x$ first,
before the death cycle, by solving the polynomial equation

\begin{equation}
\sum_i\, x^{N_i+1} = P(1-b)\,\,\,\,\,\,\,\, ,
\end{equation}

\noindent where the sum runs over all living individuals. This requirement
keeps the population constant. Equivalently, one can solve

\begin{equation}
\sum_N\, H(N)\,\, x^{N+1} = P(1-b)\,\,\,\,\,\,\,\, ,
\end{equation}

\noindent where now the sum runs over $N$ (0, 1, 2 $\dots$), and $H(N)$
counts the current number of individuals with precisely $N$ bits set along
the genome. After computing the value of $x$, we scan the whole population
($i$ = 1, 2 $\dots P$), tossing a real random number between 0 and 1 for
each individual $i$, in order to compare it with its survival probability:
if the random number is larger than $x^{N_i+1}$, individual $i$ dies.

\begin{figure}[hbt]
 \begin{center}
  \includegraphics[angle=-90,scale=0.5]{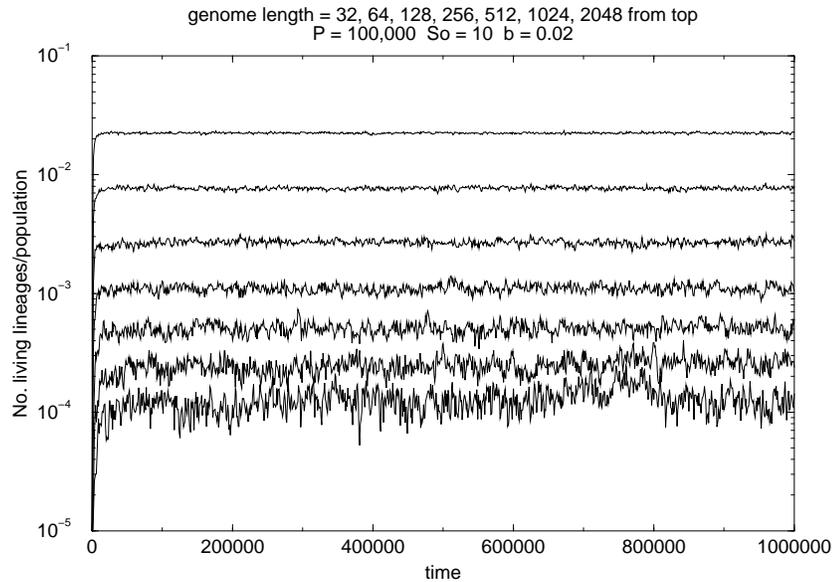}
 \end{center}
 \caption{Number of living lineages, normalised by the population, as a 
function of time, for different genome lengths.}
\end{figure}

        After each death, we choose another individual at random to be the
parent of a newborn. Its genome is copied, and some random mutations are
included at a fixed rate per bit (typically 1/32) which does not depend on
the genome length. Each mutation flips the current bit state (from 0 to 1 or
vice-versa) at a position tossed along the genome. After all mutations are
performed, the newborn is included into the population.

\begin{figure}[hbt]
 \begin{center}
  \includegraphics[angle=-90,scale=0.5]{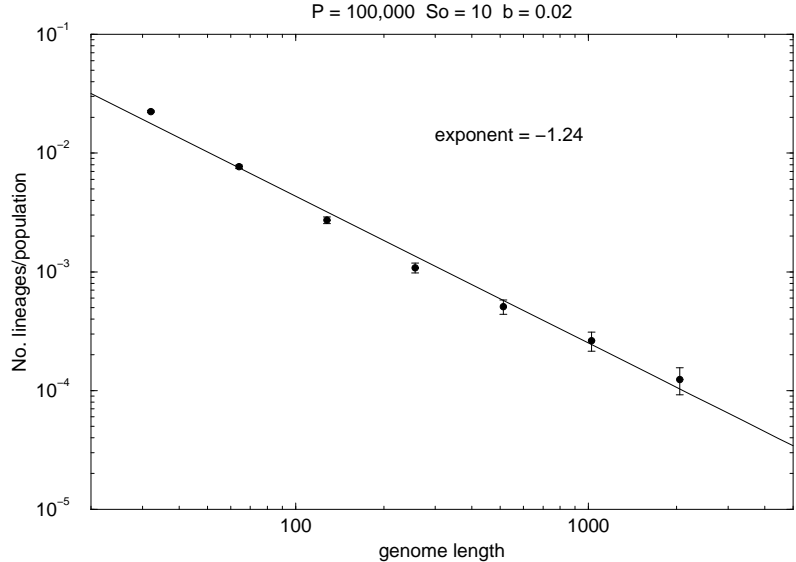}
 \end{center}
 \caption{Number of living lineages, normalised by the population, averaged 
over the final $10^5$ time steps of the simulation, as a function of the 
genome length.}
\end{figure}

        If the newborn presents fewer 1-bits than its parent, it receives
the label of potential founder of a new lineage. During the time steps that
follow, all its descendents will be monitored: if, at some future time, the
number of those descendents still alive reaches a minimum threshold $s_0$
(typically 10), then all descendents of the now confirmed founder, including
itself, are considered to belong to a new lineage.

        On the other hand, extinction occurs when the last individual of a
given lineage dies. Although a rare event, a lineage can also become extinct
if all its individuals descend from the same potential founder, being
altogether transferred to another, new, lineage, by reaching the threshold
$s_0$.

        A similar model, but without the lineage branching step, was already
used by some of us \cite{SMO03}.

\vfill\newpage\section{Results}

        We have run our program with some different sets of parameters
$\{P$, $s_0$, $b\}$. The results are qualitatively the same in all cases,
thus we will present only results for populations with $P = 10^5$
individuals, $b = 2\%$ of which die every year (immediately replaced by
newborns), requiring a minimum threshold of $s_0 = 10$ living descendents of
the same potential founder in order to have a new lineage. The genome
lengths vary from $g = 32$ up to $g = 2048$. We have also studied an
alternate version of the model in which, instead of being strictly constant,
the population is allowed to fluctuate: first, all individuals have the
chance to generate offspring, according to the rate $b$, increasing the
population; after that, the death roulette kills individuals according to
the probability $1-x^{N_i+1}$. No change is observed in what concerns the
quantities we measured below. Also, similar branching criteria were
introduced into the Penna model for biological ageing \cite{Penna,livro99},
for smaller genome lengths $g = 8$, 16, 32 and 64: the general behaviour did
not change.

        Figure 1 shows the number of living lineages as a function of time
$t$. Each time step corresponds to a scan of the whole population performing
deaths and births. We divided the number of living lineages by the constant
number of individuals, in order to show that one lineage indeed corresponds
to a considerable number of individuals (varying from approximately ten
thousand, on average, for the largest genome length of 2048 bits, down to
fifty individuals for the smallest genome length of 32 bits). One can also
observe that the total number of generations we tested, after one million
time steps, is large enough to get a stable, self-organised situation which
is, indeed, very different from the starting point, with a single lineage
and completely clean genomes.

\begin{figure}[hbt]
 \begin{center}
  \includegraphics[angle=-90,scale=0.5]{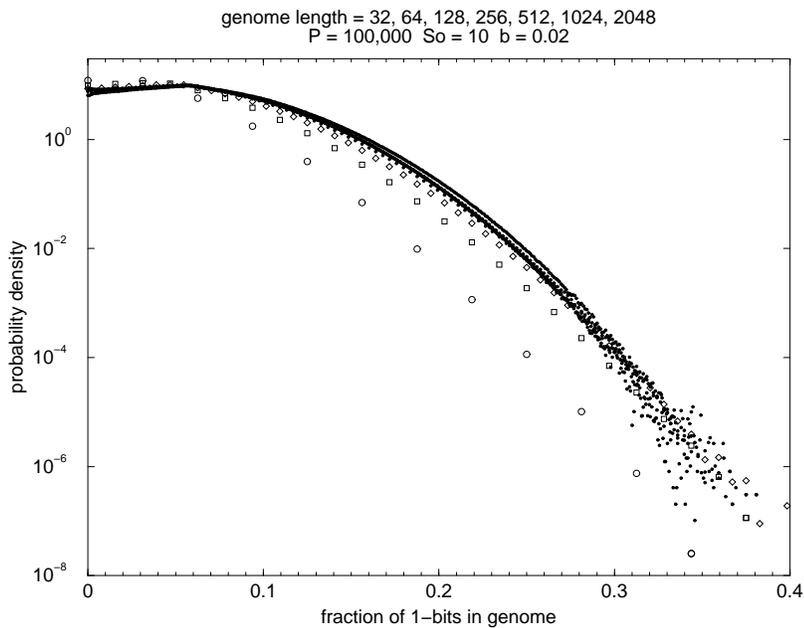}
 \end{center}
 \caption{Probability density distributions for $N/g$, where $N$ counts the
number of 1-bits along a genome of length $g$.}
\end{figure}

\begin{figure}[hbt]
 \begin{center}
  \includegraphics[angle=-90,scale=0.5]{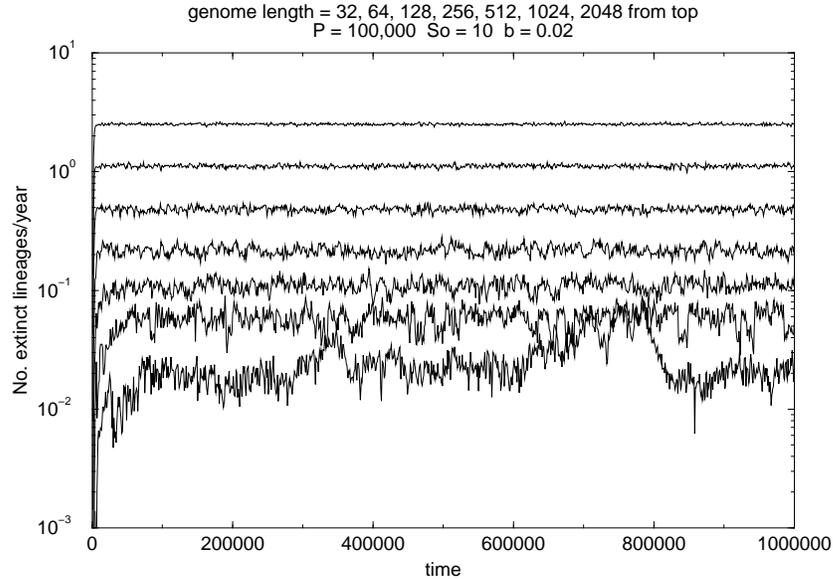}
 \end{center}
 \caption{Number of lineages which become extinct per ``year'' (one time step),
 averaged over intervals of $10^3$ time steps, as a function of time.}
\end{figure}

\begin{figure}[hbt]
 \begin{center}
  \includegraphics[angle=-90,scale=0.5]{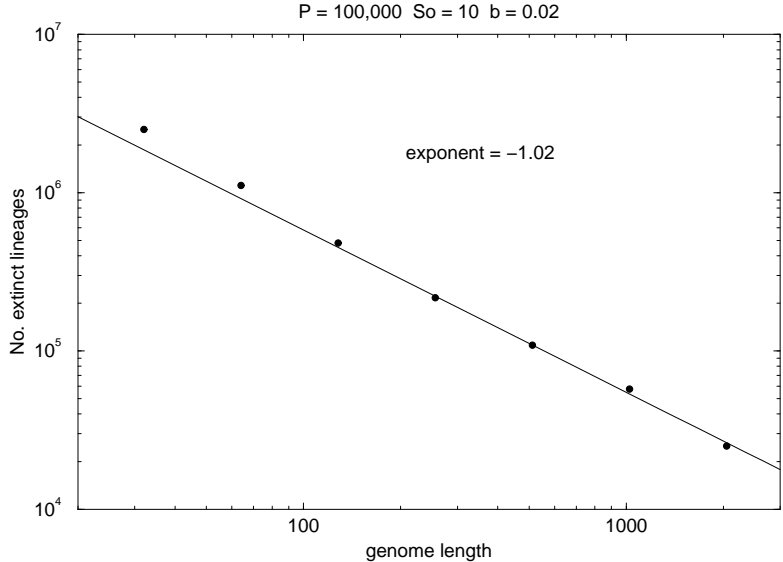}
 \end{center}
 \caption{Total number of extinct lineages as a function of the genome length.}
\end{figure}

\begin{figure}[hbt]
 \begin{center}
  \includegraphics[angle=-90,scale=0.5]{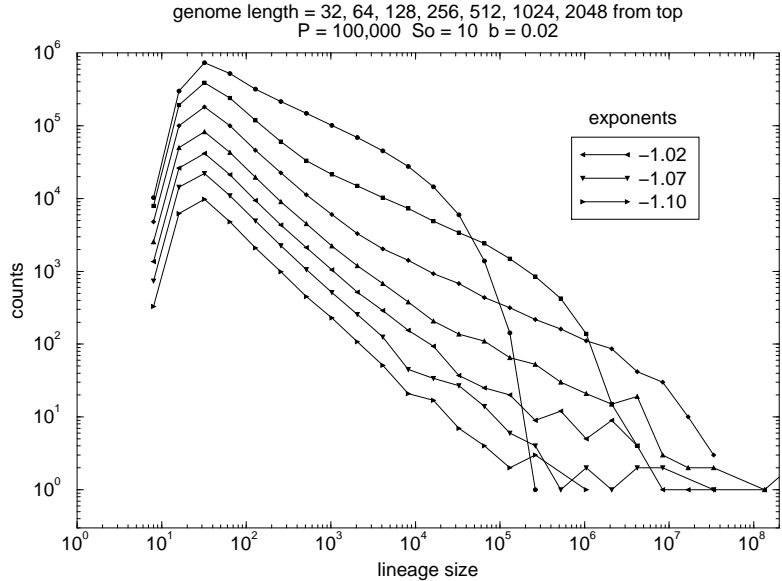}
 \end{center}
 \caption{Distribution of extinct lineages according to size $s$ (total number
 of individuals which belonged to that lineage) for different genome lengths.} 
\end{figure}

\begin{figure}[hbt]
 \begin{center}
  \includegraphics[angle=-90,scale=0.5]{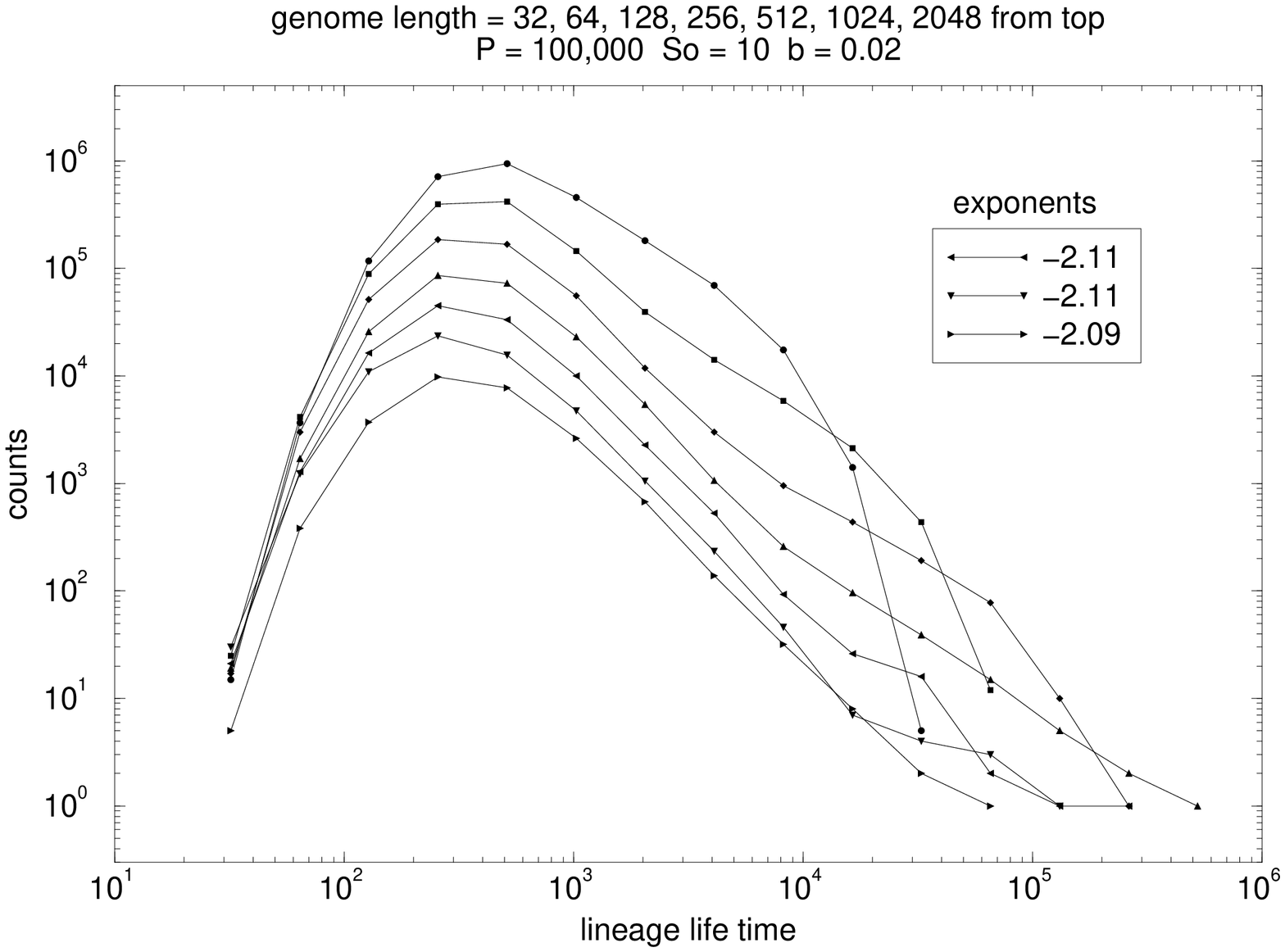}
 \end{center}
 \caption{Distribution of extinct lineages according to lifetime $\ell$, for
 different genome lengths.}
\end{figure}

        Over the last one hundred thousand time steps, after stabilisation,
we have performed the average of the number of living lineages for each
genome length. The results are displayed in figure 2. The exponent that
figures in the plot was obtained from a fit to the simulation data. For
other runs, with different sets of parameters, it remained the same. The
relation between these two quantities (number $L$ of living lineages and
genome length $g$) follows a power-law of the kind

\begin{equation}
L\, \propto\, g^{-\beta}\,\,\,\, ,\,\,\,\,\,\,\,\,
\beta \approx 5/4\,\,\,\,\,\,\,\, .
\end{equation}

\noindent Here, we propose that the numerically determined value $\beta =
1.24$ (error bar within the last digit) is in fact $\beta = 5/4$, falling
into the same family of simple multiples of 1/4, ubiquitous among biological
measurements of various kinds (see \cite{West04,West99,West98,West97,Parisi,
Demetrius} and references therein). As already quoted, G. West and
collaborators demonstrated the emergence of exponents multiple of $1/4$
based only on three fundamental ingredients. Our lineage model shares the
same ingredients, namely:

1) a multiple hierarchical branching --- in our case, lineages born from 
others;

2) a size invariant limit for the final branch --- in our case, we require a
fixed minimum population $s_0$ in order to have branching; 

3) a free-energy minimisation process --- in our case, the growing-entropy 
tendency provided by the random mutations (in the direction of randomising 
the bits along the genome as time goes by) is balanced by the selection 
mechanism (which gives preference to individuals with the smallest possible 
number of 1-bits).

        Figure 3 illustrates this last ingredient. By counting the number of
1-bits along each genome, the results are distributed far below half of the
whole length $g$ (which would be the maximum-entropy situation), showing the
efficiency of the selection process. On the other hand, the non-vanishing width
observed in the same distributions shows a high degree of genetic diversity
preserved within the survivors, even when the genome length is varied. Note
that, with the exception of the three smallest genome lengths (symbols), all
other curves (small black dots) collapse into a single,
genome-length-independent one, within the figure scale.

        Figure 4 shows the number of lineages which become extinct each
year, as a function of time. Extinction becomes more difficult for larger
genome lengths. Figure 5 shows the total number $N$ of extinct lineages,
during the whole one-million-time-step history, as a function of the genome
length. Again, we observe a power-law behaviour according to the general
trend

\begin{equation}
N\, \propto\, g^{-\gamma}\,\,\,\, ,\,\,\,\,\,\,\,\, 
\gamma \approx 1\,\,\,\,\,\,\,\, .
\end{equation}

        Figure 6 shows the distributions of extinct lineages as a function
of their sizes. One observes again a power-law behaviour with exponent very
close to 1 (even for parameters other than the ones used for this particular
plot). The exponents obtained from a fit to the data corresponding to the
largest genome lengths are shown. The position of the peak does not change
when the genome length is increased, in agreement with our criterion for
branching, namely a fixed minimum number $s_0$ of living individuals. Thus,
in the limit of large populations and large genome lengths, the probability
distribution of lineages size $P(s)$ is expected to be

\begin{eqnarray}
P(s) &=&  C\, s^{-\lambda}\,\,\,\,\,\,\,\,
{\rm if}\,\,\,\,  s \ge s_0,\,\,\,\, {\rm where}\,\,\,\, \lambda \approx 1\\
     &=&  0\,\,\,\,\,\,\,\, {\rm otherwise}\,\,\,\,\,\,\,\, .\nonumber
\end{eqnarray}

\noindent The value of $\lambda$ can be exactly 1 or slightly larger than 1,
and the constant $C$ does not depend on the genome length.

        The distribution of lineage lifetime, figure 7, is different. Its peak
position {\bf does} depend on the genome length $g$. At the same limit of large
populations and large genome lengths, its probability ${\cal P}(\ell)$ reads

\begin{eqnarray}
{\cal P}(\ell) &=&  (\alpha-1)\,\, [\ell_0(g)]^{\alpha-1}\,\, 
\ell^{-\alpha}\,\,\,\,\,\,\,\,
{\rm if}\,\,\,\, \ell \ge \ell_0(g),
\,\,\,\,\, {\rm where}\,\,\,\, \alpha \approx 2\\
     &=&  0\,\,\,\,\,\,\,\, {\rm otherwise}\,\,\,\,\,\,\,\, .\nonumber
\end{eqnarray}

\noindent Again, $\alpha$ can be exactly 2 or slightly larger than 2. This
value is in complete agreement with the real exponent found by
paleontologists John Sepkoski and David Raup from fossil data. The
multiplicative constant in front of $\ell^{-\alpha}$ can be easily obtained
by integrating equation (6) and equating the result to unity: for $\alpha =
2$, it coincides with the minimum cutoff lifetime $\ell_0(g)$ itself.

        The dependence of $\ell_0$ on $g$ also follows a power-law behaviour 

\begin{equation}
\ell_0\, \propto\, g^{-\delta}\,\,\,\, ,\,\,\,\,\,\,\,\, 
\delta \approx 1/4\,\,\,\,\,\,\,\, ,
\end{equation}

\noindent as can be seen, for instance, by plotting the peak positions on
figure 7 against $g$. Alternatively, and with better accuracy, one can plot
the average lifetime against $g$. The exponent we get from this plot (not
shown) is 0.26, for our simulational data. Indeed, a simple reasoning can
link the number $L(g)$ of living lineages at a given time, equation (3),
with the number $N(g)$ of extinct lineages during the whole history,
equation (4). The former can be counted by adding the probability of each
lineage $j$ to be alive at a given time, i.e. its lifetime $\ell_j$ divided
by the whole historical time $T$,

\begin{equation}
L(g) = \sum_{j=1}^{N(g)} {\ell_j\over T} =
{N(g)\over T} \int_{\ell_0}^T d\ell\,\, \ell\,\, {\cal P}(\ell)
\,\,\,\,\,\,\,\, .
\end{equation}

\noindent Considering $\ell_0 << T$, we get

\begin{equation}
L(g)\,\,  \propto\,\,  N(g)\,\,  \ell_0(g)\,\,\,\,\,\,\,\, ,
\end{equation}

\noindent and the consequent scaling relation

\begin{equation}
\beta = \gamma + \delta
\end{equation}

\noindent which holds in general (apart from small logarithmic corrections, if
$\alpha = 2$). This relation is very well verified by our numerical data.

        The ratio 2:1 we found between the exponents $\alpha$ and $\lambda$
governing the two probability distributions for lineages (according to their
lifetime or size) has an interesting interpretation. The growth of the number 
of lineages is not restricted by the finite size of the whole population. 
Each lineage grows by itself, reaches its maximum number of individuals, and 
then shrinks up to extinction due to its own genetic meltdown. If the 
maximum number of living individuals belonging to a lineage was somehow 
limited by an external source, then this maximum would be kept for a long 
time, waiting for the unavoidable genetic meltdown which eventually leads to 
extinction: in this case, the relation between lineage size $s$ and lifetime 
$\ell$ would be linear. On the contrary, we obtain a relation

\begin{equation}
s\, =\, A(g)\, \ell^\omega\,\,\,\, ,\,\,\,\,\,\,\,\, 
\omega \approx 2\,\,\,\, ,
\end{equation}

\noindent in agreement with the ratio $\alpha/\lambda \approx 2$ we got
previously from the lifetime and size distribution probabilities,
separately. We have measured $\omega$ independently, by accumulating a
$[\ell,s]$ histogram of all lineages, an example of which is shown in the
table, for a genome length of 64. Lineages live in a narrow stripe of the
space $[\ell,s]$, near the line defined by equation (11). For all genome
lengths, $\omega$ is always very close to 2, according to our simulational
data.

\vfill\newpage

\begin{table}
\caption{Probability distribution of species as a function of lifetime $\ell$ 
and size $s$. Fractions smaller than 0.0001 are not shown.} 
\bigskip

\begin{tabular}{|c||c c c c c c c c c c|}
\hline

        &   &   &   &   &   &   &   &   &   & \\
1048576 & . & . & . & . & . & . & . & . & . & .0001 \\
524288 & . & . & . & . & . & . & . & . & .0001 & .0001 \\
262144 & . & . & . & . & . & . & . & . & .0006 & .  \\
131072 & . & . & . & . & . & . & . & .0005 & .0007 & .  \\
65536 & . & . & . & . & . & . & . & .0018 & .0002 & .  \\
32768 & . & . & . & . & . & . & .0010 & .0020 & . & .  \\
16384 & . & . & . & . & . & . & .0035 & .0007 & . & . \\
8192 & . & . & . & . & . & .0014 & .0050 & . & . & . \\
4096 & . & . & . & . & . & .0066 & .0026 & . & . & . \\
2048 & . & . & . & . & .0010 & .0120 & .0003 & . & . & . \\
1024 & . & . & . & . & .0090 & .0102 & . & . & . & . \\
512 & . & . & . & .0003 & .0253 & .0039 & . & . & . & . \\
256 & . & . & . & .0109 & .0425 & .0008 & . & . & . & . \\
128 & . & . & .0005 & .0692 & .0372 & . & . & . & . & . \\
64 & . & . & .0323 & .1705 & .0140 & . & . & . & . & . \\
32 & . & .0105 & .2210 & .1190 & .0008 & . & . & . & . & . \\
16 & .0021 & .0644 & .1007 & .0050 & . & . & . & . & . & . \\
8 & .0015 & .0047 & .0007 & . & . & . & . & . & . & . \\
        &   &   &   &   &   &   &   &   &   & \\
\hline   
\hline
        &   &   &   &   &   &   &   &   &   & \\
$s/\ell$ & 64 & 128 & 256 & 512 & 1024 & 2048 & 4096 & 8192 & $2^{14}$ & 
$2^{15}$ \\
        &   &   &   &   &   &   &   &   &   & \\
\hline
\end{tabular}
\end{table}

        By using the identity $P(s) ds = {\cal P}(\ell) d\ell$, one can also
show the further relation

\begin{equation}
s_0 \sim A(g)\,\, [\ell_0(g)]^\omega\,\,\,\, ,\,\,\,\,\,\,\,\, {\rm or}
\,\,\,\,\,\,\,\, \ell_0(g)\,\, \propto\,\, [A(g)]^{-1/\omega}\,\,\,\,\,\,\,\, ,
\end{equation}

\noindent from which one can again (and independently) extract the exponent
$\delta$ relating $\beta$ and $\gamma$, through the proportionality constants
$A(g)$, equation (11), provided by the $[\ell,s]$-histograms.

\vfill\newpage\section{Conclusions}

        We study a simple population dynamics model where the genome of each
individual is represented by a bit-string. The survival probability
decreases with the number of 1-bits along the individual's genome. At each
time step, a certain fraction of individuals die according to these
probabilities, and are replaced by survival's offspring. The genome of each
offspring is a copy of the parent's, with a few random mutations. Lineage
branching occurs when an offspring happens to have a genome better than its
parent, provided its own descendence succeeds in growing up to surpass a
threshold of living individuals.

        By simulating this simple model on a computer, we find some general
power-law relations which seem to be {\bf independent} of the particular
parameters adopted in the simulations, and also of modifications of the
dynamic rules themselves. One of these power-laws, namely equation (6)
describing the distribution of extinct lineages per lifetime, agrees with
real paleontological data \cite{Sepkoski,Raup1,Raup2}, for which the
exponent $\alpha \approx 2$ also agrees with our numerically determined
value. No real data is available in order to compare the other exponents we
measured (equations 3, 4, 5, 7 and 11). Nevertheless, we were also able to
obtain some analytical scaling relations between these various exponents,
all of them in agreement with our numerical data. Moreover, within our
narrow error bars, all these exponents are multiple of $1/4$, in complete
agreement with the general framework theoretically studied by West {\it et
al} \cite{West99,West98,West97} in a different context. These authors show
the emergence of exponents multiple of $1/4$, which are ubiquitous within
biological systems, based only on three very general assumptions also shared
by our model. Thus, we propose these exponents could be universal, valid for
other evolutionary systems more complicated than our toy model.

\bigskip

\section*{Acknowledgements} This work is partially supported by Brazilian 
agencies CNPq and FAPERJ.

\end{document}